\documentclass{ws-fnl}
\usepackage{cite}

\newcommand{\bbZ}{\mathbb{Z}}

\begin{document}

\markboth{Schell, Sadgrove, Nakagawa, Wimberger}{(Engineering transport by concatenated maps)}

\catchline{}{}{}{}{}
%%%%%%%%%%%%%%%%%%%%%%%%%%%%%%%%%%%%%%%%%%%%%%%%%%%%%%%%%%%%%%%%%%%%

\title{Workshop on noisy many-body systems: Engineering transport by concatenated maps}

\author{Torben Schell}
\address{Institut f\"u{r} Theoretische Physik, Universit\"{a}t Heidelberg, Philosophenweg 19, 69120 Heidelberg, Germany}
\author{Mark Sadgrove}
\address{Institute for Laser Science, The University of Electro-Communications, Chofushi, Chofugaoka 1-5-1, Japan}
\author{Ken'ichi Nakagawa}
\address{Institute for Laser Science, The University of Electro-Communications, Chofushi, Chofugaoka 1-5-1, Japan}
\author{Sandro Wimberger}
\address{Institut f\"u{r} Theoretische Physik and Center for Quantum Dynamics, Universit\"{a}t Heidelberg, Philosophenweg 19, 69120 Heidelberg, Germany \\
s.wimberger@thphys.uni-heidelberg.de}

\maketitle

\begin{history}
\received{(\today)}
%\revised{(revised date)}
%\accepted{(Day Month Year)}
%\comby{(xxxxxxxxxx)}
\end{history}

\begin{abstract}
We present a generalized kick rotor model in which the phase of the kick can vary from kick to kick. This additional freedom allows one to 
control the transport in phase space. For a specific choice of kick-to-kick phases, we predict novel forms of accelerator modes which are 
potentially of high relevance for future experimental studies.
\end{abstract}

%\pacs{05.45.Mt, 03.75.-b, 05.60.Gg}

\section{Motivation}
\label{moti}

For the standard map (often also called Chirikov-Taylor map) there is hardly any introduction necessary. It is one of the standard models
of low-dimensional Hamiltonian chaos theory \cite{LL}. Also its quantized version, usually known as the quantum kicked rotor, has a long history of
research \cite{casati}, see also for instance \cite{Izr,Fish,raizen} for older and \cite{SW} for a more recent review. The quantum kicked rotor
has been of special interest over the last three decades because of two dynamical regimes, one which supports ballistic motion for 
resonant driving \cite{Izr,SW}, and a second one where transport is suppressed in the same way as in an Anderson impurity model \cite{casati,Fish,loc}.

Many variants of the quantum kicked rotor are known, including realizations of quantum ratchets \cite{ratchet,sad-rat,Gil-2,Gil-3} and kicking in the
presence of an additional gravity field \cite{acc}, to name just a few. The kicking along the direction of gravity introduces new accelerator
modes supporting directed ballistic transport \cite{FGR}. A mechanism for fast transport was recently also proposed in reference \cite{italo}. Here we 
propose an interesting alternative to engineer the kick-to-kick phase of the potential, which may turn out to be very useful for future 
experimental applications \cite{SSNW}.

Phase noise in general tends to destroy quantum mechanical effects, thus hindering transport close to the above mentioned quantum resonant
driving \cite{sad-08} for instance. Yet, in other contexts, a noisy environment may even help to enhance the signal, a famous example being
stochastic resonance in classical \cite{sr} and quantum systems \cite{q-sr}, or the enhancement of Landau-Zener tunneling in noise-driven
lattices \cite{TMW}. Here we study the impact of {\it controlled} phase jumps of a periodic potential which imparts the kicks. In the limit of
an infinite sequence (or alphabet) of jumping values, this corresponds essentially to a phase randomization, whilst for a finite set, the classical and the quantum kicked dynamics can be described by concatenated maps. The latter form a new map with a new temporal period for the evolution of the kicked system.

In this paper, we analyze in the next section a specific classical evolution based on a realization of phase jumps with a small alphabet,
implying a new map with a period of four single kicks. We study the fixed points of this new map and their transporting behavior. Section \ref{exp}
discusses implications of the classical evolution on a quantum realization based on cold atoms kicked with a standing wave whose relative phase
in position space is changed from kick to kick. We first use a slightly amended classical (so-called pseudo-classical) model to describe
approximately the quantum evolution. Then we show how our system proves useful for the population of the predicted transporting islands by 
initial quantum states anchored to them. Section \ref{sum} concludes the paper.

\section{The classical kicked rotor and phase jumps}
\label{class}

\subsection{The general model}
\label{model}

Our system is described by the classical kicked-rotor Hamiltonian:
\begin{equation}
\label{eq:1}
H(t') = \frac{p^2}{2} + K  \cos(\theta + \phi_{t'}) \sum_{t\in \bbZ} \delta (t'-t) \,,
\end{equation}
with the additional time-dependent phase $\phi_{t'}$ in the sinusoidal potential, as compared to the original model \cite{LL}.
Since the kicks act instantaneously, the potential is zero in the moments of time when no kick acts. This means that our phase
term is only important when the kicks are flashed on, i.e. we can equally set $\phi_{t'}=\phi_{t}$. Moreover, the instantaneous 
kicks (imparted with strength $K$) allow for a rather simple dynamical description of the system by a mapping from just before one kick to just 
before the next one \cite{LL}, of the form:
\begin{eqnarray}
p_{t+1} & = & p_{t} + K \sin(\theta_t + \phi_t) \nonumber \\
\theta_{t+1} & = & \theta_{t} + p_{t+1} ~ {\rm mod} ~ 2 \pi \,.
\label{eq:2}
\end{eqnarray}
For the standard map $\phi_{t}=const.=0$, and hence the map is periodic with the periodicity of one kick. Now if we allow $\phi_{t}$ to be chosen
from a finite set and fix the order of their appearance, the map remains periodic, just with a longer period given by the size of the possible choices
for the phase (or the size of the alphabet). This is the case we would like to study in the following in more detail. The central question is how we 
can control the transport in momentum space by a proper choice of the sequence of the values $\phi_{t}$. 

\subsection{Case study}
\label{ex}

\begin{figure}[t]
\centerline{\psfig{file=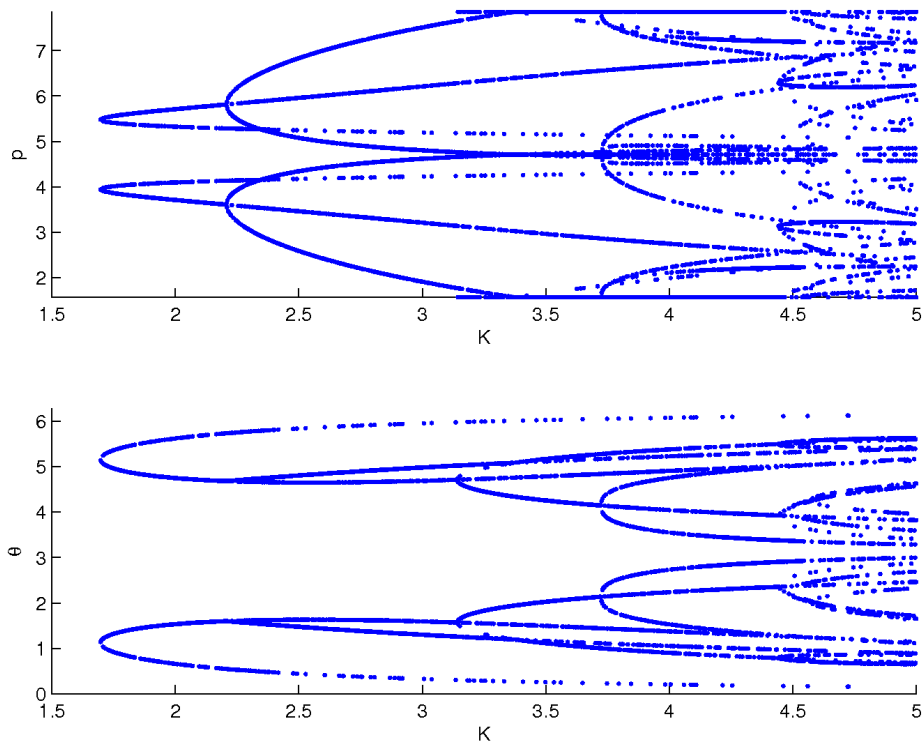,width=12.5cm}}
\vspace*{8pt}
\caption{Momentum (a) and angle coordinate (b) of fixed points with the property that $M_{\rm 4}$ maps $(p,\theta)$ into $(p\pm 2\pi, \theta + 10\pi ~{\rm mod}~(2\pi) = \theta)$,
found as a function of the kick strength $K$. We observe a series of bifurcations at which certain fixed points become unstable but producing new stable ones 
(with smaller stability islands around them) at certain values of $K$.}
\label{fig:1}
\end{figure}

To keep the description simple we focus from now on a specific choice of phases, which follows the pattern given by $\{0,\pi/2,0,-\pi/2,\ldots\}$. 
A posteriori this choice will be justified by the interesting transporting islands it gives in the new phase space induced by the new periodic map obtained by applying
four times Eq.~(\ref{eq:2}). Each of these four steps may be represented by a nonlinear function from the initial to the final two variables $(p_t,\theta_t) \to 
(p_{t+1},\theta_{t+1})$. We denote $M_{\phi}$, with $\phi = 0, \pi/2, - \pi/2$ for the first, second and fourth step respectively. 
The new time-periodic map is then just the concatenation of these functions:
\begin{equation}
M_{\rm 4} = M_{-\pi/2} \circ M_0 \circ M_{\pi/2} \circ M_0 \,.
\label{eq:3}
\end{equation}
The classical evolution is visualized by iterating the map $M_{\rm 4}$ and plotting the trajectories in Poincar\'{e} surface of sections \cite{LL}.
In order to find such islands of local stability in the Poincar\'{e} maps, the task is the identify fixed points of the {\it new} map $M_{\rm 4}$. 
In the limit of small $K$, one can easily find the following stable fixed points in ($p,\theta$): $(\pi,\pi), (\pi/2,0), (3\pi/2,\pi), (0,\pi)$.
All of them are surrounded by relatively large stable islands in phase space (for simplicity we call Poincar\'{e} surface of sections just phase spaces from now on).
All have unstable partner fixed points (by the Poincar\'{e}-Birkhoff theorem \cite{LL}), but the stable islands also disappear as $K$ grows at some point. As
one may easily verify, for all these fixed points, the momentum variable indeed returns to the initial value. In other words, all these islands are static and 
non-transporting in momentum space. That is the reason for looking at more interesting parameter values of larger $K$ in the following.

Because of the nonlinearity of the problem, we obtain fixed point equations which are quite non-transparent, and for general values of the single parameter $K$, analytical solutions are not simple. We may proceed instead by solving the fixed point equations numerically. 
Once identified we can search for them in phase space. There an island should be visible if the fixed point is stable, thus the stability is 
automatically checked in this way. Alternatively, one may numerically check the stability of the found fixed points by studying the stability of the 
linearized map close to them \cite{LL}. 
Figure \ref{fig:1} shows the numerically found fixed points of a special class (see caption) as a function of the kicking strength $K$ of the new map. 
At $K \approx 1.7$, two branches are seen, which for increasing kick strength bifurcate more and more into an entire series of fixed points. The most prominent stable fixed points are
characterized by the largest size of the island in phase space surrounding them. The area of these structures is indeed important, 
once we want to check the relevance of our findings for the quantum version of our map $M_{\rm 4}$, as we will do in the following section.

\begin{figure}[t]
\centerline{\psfig{file=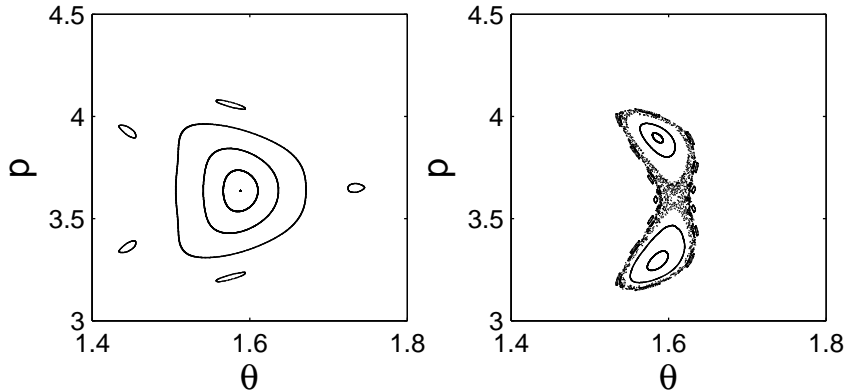,width=11.5cm}}
\vspace*{8pt}
\caption{Poincar\'{e} surface of sections for $K=2.154$ (left) and $K=2.24$ (right), just below and above the bifurcation of the central island shown in the left panel. 
At $K \approx 2.2$, this islands splits into two new ones corresponding to two new stable fixed points. The new islands are typically smaller, hence the case
shown in the left panel is better for applications in the quantum case, where the area has to be compared to the size of a Planck's cell.
}
\label{fig:2}
\end{figure}

Figure \ref{fig:2} presents the islands of stability for two different values of the kick strength just before and after a bifurcation as shown in the previous figure around 
$K \approx 2.2$.
These Poincar\'{e} surface of sections are obtained by taking also the modulus of the momentum variable in all the applied iterations of the map of Eq.~(\ref{eq:3}). Please note that
all phase spaces are periodic in momentum with period $2\pi$ as well, a trivial consequence of the sinusoidal kicking potential. Hence, instead of jumping from one phase space cell
to the next one during the evolution given by $M_{\rm 4}$, we let the trajectories form regular islands iterating over many applications of $M_{\rm 4}$. That the islands are indeed 
transporting, i.e. the momentum (for any initial state chosen within the regular islands) increases linearly with the number of kicks, is easily seen by not taking the modulus
in momentum space. We refrain, however, to show such data here, since we will present a similar plot in comparison with the quantum case, see figure \ref{fig:3} in the next section.

Before we discuss possible experimental observations of the new accelerator modes, we conclude with a comment on the global phase space structure. The reader may easily convince him/herself that the modified standard map of Eq.~(\ref{eq:3}) obeys an inversion symmetry around the point $(\theta=\pi, p=3\pi/2)$. This symmetry implies the pair-wise existence of accelerator modes, one flying in the direction of lower momenta $p$, the second one in the direction of larger $p$. In other words, the global net transfer of momentum is zero, reminiscent of the sum rules for Hamiltonian ratchet systems \cite{ratchet,ketz}. Finally, as long as the kick strength is sufficiently large, i.e. $K > 2$, the islands corresponding to the new modes are surrounded by a chaotic sea. The latter is not visibly affected by the additional modulation of the potential phase, meaning that we see normal diffusion with and without the modulation.

\section{Experimental implications}
\label{exp}

Apart from very recent realizations of kicked rotor dynamics by pulses propagating across optical fibers \cite{ulf}, the standard technique has been pioneered by
Raizen and co-workers \cite{raizen} using cold atoms. Nowadays experiments are performed with Bose-Einstein condensates, 
which allow for a much better control of the initial conditions of the atoms in phase space, see e.g. \cite{Gil-1,Gil-2,Gil-3,NZ,sad-rat} 
for some recent experiments. Here the atoms always move along a line while kicked by a standing wave of laser light. 
Since the atoms are moving along a one-dimensional line (neglecting other dimensions whose influence is indeed minimal in 
non-interacting atomic systems), the kicked rotor model has to be slightly amended. This can be done by reducing the quantum evolution in a periodic potential to the 
motion along a circle using Bloch's theorem \cite{WGF}. Then the experimental coordinates for the atomic center of mass momentum $P$ (in units of 2-photon recoils of the 
standing wave) and the atomic position $X$ (in units of the spatial period of the standing wave) transform into $P=n+\beta$ and $\theta= X~{\rm mod}~2\pi$, where
the integer $n$ is the angular momentum and $\beta \in [0,1)$ the quasi-momentum. By Bloch's theorem the quasi-momentum is a constant of the motion, and in practice
it just determines the initial position of the atoms along the momentum axis in classical phase space \cite{SW,SW-05}.

The evolution defined by the map of Eq.~(\ref{eq:2}) changes now in view of the experimental situation to \cite{FGR,WGF,SW,SW-05,sad-prl}
\begin{eqnarray}
I_{t+1} & = & I_{t} + K \sin(\theta_t + \phi_t) \nonumber \\
\theta_{t+1} & = & \theta_{t} + I_{t+1} ~ {\rm mod} ~ 2 \pi \,,
\label{eq:4}
\end{eqnarray}
with
\begin{equation}
I_{t}  =  p_t + \ell \pi + \tau \beta \,.
\label{eq:5}
\end{equation}
Here $\ell \in \{0,1,2,...\}$ is integer (known as the order of quantum resonant motion) and $\tau$ is the dimensionless kicking period for the original model.
In experimental parameters it is given by $\tau = (T/T_T)4\pi$, where $T$ is the same period in seconds and the so-called Talbot time is $T_T \approx 66.3 \rm \mu s$ 
for $^{87}$Rb atoms \cite{SW,Gil-1,Gil-3} for instance. The classical kicking parameter $K$ must be identified as
$K = k \tau $ for $\ell=0$ and $K=k\epsilon$ for $\tau=2\pi\ell+\epsilon$ (where we restrict to positive detunings $\epsilon$ from the 
quantum resonant values $\tau=2\pi\ell$ \cite{Izr,SW} for notational simplicity), with the dimensionless kicking strength $k$ proportional to the intensity of the
standing wave, see for instance \cite{SW,sad-prl} for details.

The classical map Eq.~(\ref{eq:4}) is a good approximation for the true quantum evolution of kicked atoms as long as either $\tau$ (for $\ell=0$) or $\epsilon$ 
(for $\ell>0$) is small. Then they represent a limit of the quantum motion which is known as $\epsilon$-classical or pseudo-classical limit \cite{FGR,WGF,SW}.
Our concatenated map Eq.~(\ref{eq:3}) is formed just by iterating Eq.~(\ref{eq:4}) four times with the sequence of phases used, e.g., in section \ref{ex}. Hence,
the form of Eq.~(\ref{eq:2}) remains formally unaltered, and we can now compare the prediction of the previous section with full quantum computations. For the
latter we have to introduce the quantized version of (\ref{eq:4}):
\begin{equation}
{\hat U}_{\rm 4} = {\hat U}_{-\pi/2} \circ {\hat U}_0 \circ {\hat U}_{\pi/2} \circ {\hat U}_0 \,,
\label{eq:6}
\end{equation}
where the hat on the right side of the equation denotes the single kick quantum operators which are defined by
\begin{equation}
{\hat U}_{\phi_t} = e^{- i \tau P^2/2} e^{- i k \cos ({\hat \theta}+\phi_t)} = e^{- i \tau (n+\beta)^2/2} e^{- ik \cos ({\hat \theta}+\phi_t)}\,.
\label{eq:7}
\end{equation}

\begin{figure}[t]
\centerline{\psfig{file=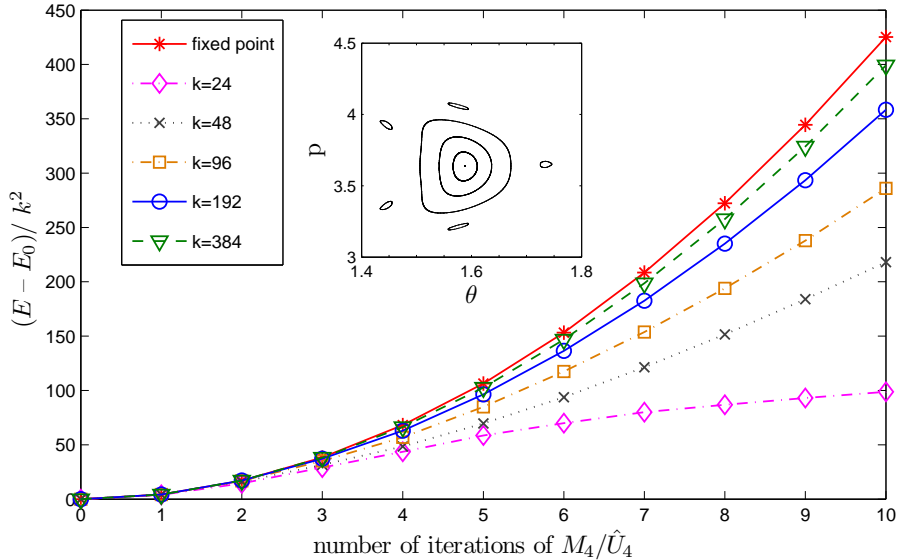,width=12cm}}
\vspace*{8pt}
\caption{Rescaled energy of a quantum wave packet anchored to the transporting island from figure \ref{fig:2} (repeated in the inset) 
as a function of iterations of the maps ${\hat U}_{\rm 4}$ or $M_{\rm 4}$. $E_0$ is the initial energy of the wave packet. 
Data are shown for increasing $k$ and decreasing $\epsilon$ at $k\epsilon=K=2.154$. The smaller $\epsilon$, the closer 
the quantum result is to the pseudo-classical evolution at the fixed point corresponding to the traveling island 
(stars, most upper line).
}
\label{fig:3}
\end{figure}

The central result of this paper is that the transporting islands in phase space can indeed be populated by quantum wave packets anchored to them. The evolution defined 
by the new quantum map Eq.~(\ref{eq:6}) then allows one to produce ballistically accelerated atoms. A simple way to check this is to plot the average energy of the
corresponding quantum wave packets as a function of the number of iterations of Eq.~(\ref{eq:6}). Figure \ref{fig:3} nicely shows the quadratic increase of the 
energy expectation value for appropriate parameters. The best increase can be reached in the limits $\tau \to 0$ or $\epsilon \to 0$ where the pseudo-classical approximation
of the quantum motion is most accurate. Therefore, the smallest value of $\epsilon$ in figure \ref{fig:3} closely follows the classical result (obtained by evolving a 
classical ensemble of initial points in phase space within the resonance island from the left panel in figure \ref{fig:2}). The size of a quantum mechanical state, 
the Planck cell, is in the pseudo-classical approximation determined by $2\pi\tau$ or $2\pi\epsilon$ respectively \cite{SW,SW-05,sad-prl}, explaining the good correspondence 
in this limits. 

\section{Summary}
\label{sum}

We presented a proposal for realizing interesting transport phenomena with generalized kicked rotor evolutions. 
Allowing for specific sequences of time-dependent phases in the kicking potential we analyzed the classical dynamics
of the corresponding generalized mappings. In particular, we report the existence of traveling islands in phase space,
which turn out to be useful also for the quantum realization of the model with cold atoms \cite{SSNW}. State-of-the-art 
experiments with Bose-Einstein condensates could populate the predicted islands by controlling the 
initial conditions in phase space very well \cite{Gil-1,SW}. The higher the kicking strength (and the smaller the effective
Planck's constant) can be made, the better one will observe ballistic motion of those initial wave packets.

The theoretical task remains to generalize our findings to larger alphabets of possible choices for the phase jumps. These
may produce larger sizes of stable islands, which would be easier to control experimentally in the deeper quantum regime.
The new ballistic modes would then form an optimal setting to detect chaos-- or resonance-assisted tunneling \cite{tunneling} 
of the quantum wave packets out of the classical island structure.

\section*{Acknowledgments}

It is our pleasure to thank R\`emy Dubertrand for helpful discussions, and Italo Guarneri and Shmuel Fishman for useful suggestions during the 
workshop on {\it Noisy many-body systems} in Heidelberg. S.W. acknowledges financial support from the DFG (FOR760), the Helmholtz Alliance 
Program EMMI (HA-216), the HGSFP (GSC 129/1), and the Enable Fund from the Faculty of Physics and Astronomy at Heidelberg University.


\begin{thebibliography}{0}

\bibitem{LL}
B. V. Chirikov, Phys. Rep. {\bf 52}, 263 (1979);
A. J. Lichtenberg and M. A. Lieberman, {\it Regular and Chaotic Dynamics} (Springer, Berlin, 1992).

\bibitem{casati}
G. Casati, B. Chirikov, J. Ford, F. Izrailev  in {\em {Stochastic Behavior in Classical and Quantum Hamiltonian Systems}},
ed. by G. Casati and J. Ford (Springer, Berlin, 1979), p.\ 334.

\bibitem{Izr} 
F.M. Izrailev, Phys. Rep. \textbf{196}, 299 (1990). 

\bibitem{Fish}
S. Fishman, in Quantum Chaos, School ``E. Fermi'' CXIX, ed. by G. Casati, I. Guarneri, and U. Smilansky (IOS, Amsterdam, 1993).

\bibitem{raizen} 
F. L. Moore, J. C. Robinson, C. F. Bharucha, Bala Sundaram, and M. G. Raizen, Phys. Rev. Lett. {\bf 75}, 4598 (1995);
M. G. Raizen, Adv. At. Mol. Phys. {\bf 41}, 43 (1999). 

\bibitem{SW}
M. Sadgrove and S. Wimberger, Adv. At. Mol. Opt. Phys. {\bf 60}, 315 (2011).

\bibitem{loc} S. Fishman, D. R. Grempel, and R. E. Prange, Phys. Rev. Lett. \textbf{49}, 509 (1982).

\bibitem{Gil-2}
I. Dana, V. Ramareddy, I. Talukdar, and G. S. Summy, Phys. Rev. Lett. {\bf 100}, 024103 (2008).

\bibitem{Gil-3}
R. K. Shrestha, J. Ni, W. K. Lam, S. Wimberger, and G. S. Summy, Phys. Rev. A {\bf 86}, 043617 (2012).

\bibitem{sad-rat}
M. Sadgrove, M. Horikoshi, T. Sekimura, and K. Nakagawa, Phys. Rev. Lett. {\bf 99}, 043002 (2007)

\bibitem{ratchet}
P. H. Jones, M. Goonasekera, D. R. Meacher, T. Jonckheere, and T. S. Monteiro, Phys. Rev. Lett. {\bf 98}, 073002.

\bibitem{acc} 
M. K. Oberthaler, R. M. Godun, M. B. d'Arcy, G. S. Summy, and K. Burnett, Phys. Rev. Lett. \textbf{83}, 4447 (1999).

\bibitem{FGR}
S. Fishman, I. Guarneri, and L. Rebuzzini, J. Stat. Phys. {\bf 110}, 911 (2003).

\bibitem{italo}
J. Wang, I. Guarneri, G. Casati, and J. Gong, Phys. Rev. Lett. {\bf 107}, 234104 (2011).

\bibitem{SSNW}
M. Sadgrove, T. Schell, K. Nakagawa, and S. Wimberger, Phys. Rev. A {\bf 87}, 013631 (2013).

\bibitem{sad-08}
M. Sadgrove, S. Wimberger, S. Parkins, and R. Leonhardt, Phys. Rev. E {\bf 78}, 025206(R) (2008).

\bibitem{sr} 
L.~Gammaitoni, P. H\"anggi, P. Jung, and F. Marchesoni, Rev. Mod. Phys. {\bf 70}, 223 (1998);
M. I. Dykman and P. V. E. McClintock, Nature (London) {\bf 391}, 344 (1998).

\bibitem{q-sr}
T.~Wellens, V. Shatokhin, and A. Buchleitner, Rep. Prog. Phys. {\bf 67}, 45 (2004);
D. Witthaut, F. Trimborn, and S. Wimberger, Phys. Rev. Lett. {\bf 101}, 200402 (2008).

\bibitem{TMW}
G. Tayebirad, R. Mannella, and S. Wimberger, Phys. Rev. A {\bf 84}, 031605(R) (2011).

\bibitem{ketz}
H. Schanz, M.-F. Otto, R. Ketzmerick, and T. Dittrich, Phys. Rev. Lett. {\bf 87}, 070601 (2001).

\bibitem{ulf} 
Ch. Bersch and U. Peschel, private communication.

\bibitem{NZ}
A. Ullah and M. D. Hoogerland, Phys. Rev. E {\bf 83}, 046218 (2011).

\bibitem{WGF} 
S. Wimberger, I. Guarneri, S. Fishman, Phys. Rev. Lett. \textbf{92}, 084102 (2004); 
Nonlinearity \textbf{16}, 1381 (2003).

\bibitem{SW-05}
M. Sadgrove and S. Wimberger, J. Phys. A: Math. Gen. {\bf 38}, 10549 (2005).

\bibitem{sad-prl} 
M. Sadgrove, S. Wimberger, A. S. Parkins, R. Leonhardt, Phys. Rev. Lett. \textbf{94}, 174103 (2005);
S. Wimberger, M. Sadgrove, A. S. Parkins, R. Leonhardt, Phys. Rev. A \textbf{71}, 053404 (2005).

\bibitem{Gil-1}
G. Behinaein, V. Ramareddy, P. Ahmadi, and G. S. Summy, Phys. Rev. Lett. \textbf{97}, 244101 (2006);
I. Talukdar, R. Shrestha, and G. S. Summy, Phys. Rev. Lett. \textbf{105}, 054103 (2010).

\bibitem{tunneling}
M. Sheinman, S. Fishman, I. Guarneri, and L. Rebuzzini, Phys. Rev. A 73, 052110 (2006);
S. Keshavamurthy and P. Schlagheck (Eds.), {\it Dynamical Tunneling: Theory and Experiment},  
CRC Press, Taylor and Francis Group, Boca Raton (2011).

\end{thebibliography}
\end{document}